\documentclass[9pt, twocolumn]{article}
\usepackage[utf8]{inputenc}
\usepackage[english]{babel}
\usepackage{amssymb}
\usepackage{amsmath}

\usepackage{lineno}

\usepackage[margin=0.75in, top=0.75in]{geometry}

\usepackage{algorithm}
\usepackage[noend]{algpseudocode}

\usepackage{float}

\floatstyle{boxed}
\newfloat{deffloat}{h}{deffloats}
\floatname{deffloat}{Definitions}

\usepackage{framed} 
\usepackage[framed]{ntheorem}
\theoremstyle{nonumberplain}
\newframedtheorem{frm-thm}{}

\usepackage{multicol}
\usepackage{sectsty}
\sectionfont{\bfseries\Large\raggedright}
\sectionfont{\bfseries\large\raggedright}

\usepackage{enumitem}
\newlist{todolist}{itemize}{2}
\setlist[todolist]{label=$\square$}

\usepackage{xifthen}
\newcommand{\E}[2][]{%
  \ifthenelse{\isempty{#1}}{%
    \mathrm{E}\left[#2\right]
  }{%
    \mathrm{E}\left[#2 \middle| #1\right]
  }%
}
\newcommand{\Var}[1]{\mathrm{Var}\left(#1\right)}
\newcommand{\Cov}[2][]{%
  \ifthenelse{\isempty{#1}}{%
    \mathrm{Cov}\left(#2\right)
  }{%
    \mathrm{Cov}\left(#2, #1\right)
  }%
}

\usepackage{graphicx}
\graphicspath{ {./images/} }
\usepackage[export]{adjustbox}
\usepackage{subcaption}

\usepackage[
backend=biber,
style=alphabetic,
citestyle=authoryear
]{biblatex}
\usepackage[colorlinks,citecolor=red,urlcolor=blue,bookmarks=false,hypertexnames=true]{hyperref}

\author{
  Jeffrey Wong\\
  Experimentation Platform\\
  Netflix, Inc.
  \and
  Randall Lewis\thanks{Lewis was Director of Economics at Netflix, Inc when this work completed. He is now Chief Scientist at Nanigans, Inc.}\\
  Science and Analytics\\
  Netflix, Inc.
  \and
  Matthew Wardrop\\
  Experimentation Platform\\
  Netflix, Inc.
}

\title{Efficient Computation of Linear Model Treatment Effects in an Experimentation Platform}

\date{September 2019}

\begin{document}
\maketitle

\begin{abstract}
    Linear models are a core component for statistical software that analyzes treatment effects. They are used in experimentation platforms where analysis is automated, as well as scientific studies where analysis is done locally and manually. However, the implementations of linear models for online platforms and for local analysis are often different due to differing demands in scalability. To improve leverage, we developed one library that powers both modes of analysis, consolidates many forms of treatment effects from linear models, and optimizes the computational performance of the software.
    The highly performant algorithms allow the library to run linear models at Netflix's scale and reply to live queries for treatment effects from an online service without using distributed systems in seconds. Wrapping these algorithms in a library that can be used interactively make large scale linear models accessible for both online platforms and scientists to use locally.
\end{abstract}

\section{Introduction}

Causal effects have been studied in the context of randomized and controlled experiments, which may be orchestrated by an experimentation platform (XP). Much literature has been dedicated to the engineering and management of multiple simultaneous experiments by an XP, for example \cite{gupta2018anatomy}, as well as the outcomes of those experiments, such as \cite{kohavi2012trustworthy}. However, there are only few references on how an engineering platform analyzes experiments, namely the analytical demands of an XP (\cite{xp}), CUPED (\cite{deng2013improving}) and FORME (\cite{guo2015flexible}). Many experimentation platforms have standardized on the statistical model, CUPED, which has been adopted by at least Microsoft, Uber, and Booking.com (\cite{uber} and \cite{booking}) for its impact and easy implementation. Meanwhile, in other causal inference fields, many new models have been developed to measure different forms of causal effects and with greater robustness, for instance using causal random forests and instrumental variables (\cite{wager2018estimation}, \cite{angrist1996identification}). However, adoption of these methods into an XP are difficult because of challenges in generalizing and scaling the methods into a platform. 

To motivate the demands for scalability, we describe the user experience using Netflix's XP. First, a user logs into the XP and selects an experiment to analyze. The XP retrieves data from a warehouse corresponding to the experiment and runs live statistical tests. Results and visualizations are displayed in a dashboard. The user can interact with the dashboard and analyze different views of the data, which will trigger the XP to pull new data, run statistical tests, and render visualizations. This interactivity is the motivator for an engineering system that can scale and respond to live queries.

Netflix created a focus on Computational Causal Inference (CompCI) in order to address scalability challenges. This has resulted in the development of a high performance framework for computing treatment effects. We began with treatment effects from linear models due to their simplicity, flexibility and power. Hereafter, we refer to treatment effects from linear models as linear model treatment effects.

Linear models provide a flexible framework that can improve statistical power over simple two-sample t and proportions tests, as well as estimate a full variety of treatment effects that are relevant to XPs (\cite{xp}) such as:
\begin{enumerate}
    \item The average treatment effect (ATE), which is used to summarize the performance of the treatment overall.
    \item The conditional average treatment effect (CATE), which measures the treatment effect for a specific subpopulation.
    \item The dynamic treatment effect (DTE), which measures the treatment effect through time and is a core element to forecasting the effect after the experiment concludes.
\end{enumerate}
In fact, both CUPED and FORME are subclasses of linear models.
To have a general purpose linear model framework for all variants of the treatment effect and for any form of the linear model, we leverage the potential outcomes framework first used by Neyman (\cite{splawa1990application}) and summarized in \cite{ rubin_potentialoutcomes}, which describes the estimation of treatment effects as a counterfactual scoring problem. 
We use this single framework to optimize the computational strategy for both fitting linear models and evaluating treatment effects. At Netflix, the strategy is fast enough to run in an online and interactive web service, responding to live analysis queries in seconds, without precomputing results. It is also general and future-proof so that causal inference researchers have freedom to iterate on the class of models, the form of the models, and the form of the variance. Finally, our implementation can run on a single machine, demonstrating a path that has little engineering overhead to scale a causal effects model into an XP. We use Ordinary Least Squares (OLS) as a motivating example throughout the paper, though the computational strategy is easily extended to other linear models such as quantile regression.

\section{Counterfactual Scoring Framework}

In randomized and controlled experiments we aim to answer a question: what would have happened if we applied the treatment experience to the users? This hypothetical, also known as the counterfactual, cannot be observed because a user cannot simultaneously experience both the control and the treatment. Instead, the random assignment mechanism from the experiment allows researchers to infer the counterfactual outcome, and thus the expected difference in an outcome caused by the treatment. 
Many statistical tests can be performed within a counterfactual scoring framework.
Adopting this framework provides an experimentation platform with a single and well-defined computational strategy for measuring different treatment effects, such as the ATE, CATE, and DTE. This standardization is important for developing optimizations targeted for the XP. The remainder of this paper illustrates how OLS with counterfactual scoring can provide rich measurements on causal effects in an efficient and statistically robust manner.

\subsection{Average Treatment Effect in OLS}

Suppose we have a randomized and controlled experiment in which users experience one of: the control experience, denoted $X = 0$, or the treatment experience, $X = 1$. If the outcome of interest is $y$, then the average treatment effect, ATE, is defined as
$$ATE = \E[X = 1]{y} - \E[X = 0]{y}$$ which is a difference in counterfactual scores.
The standard way of estimating the ATE given a sample of $n$ users, divided into $n_C$ users in the control group and $n_T$ in the treatment group, is an empirical difference:
$$\widehat{ATE} = \frac{1}{n_T} \sum_{i: X_i = 1} y_{i} - \frac{1}{n_C} \sum_{i: X_i = 0} y_{i}.$$
This difference in averages is the effect size reported by the common two-sample t-test, which is used by many experimentation platforms to test if the means of the control and treatment groups differ. 

These two methods, estimating the difference in sample averages, and estimating the expected counterfactual outcome from a model, need not be distinct. The two-sample t-test is equivalent to the Simple OLS, and can be thought of as a model that estimates counterfactual scores.

\textbf{Claim}: Assume a linear model with the form $y = \alpha + X\beta_1 + \epsilon$, where $\alpha$ is an intercept, $X$ is a binary indicator variable with $X=1$ denoting the treatment experience and $X=0$ the control experience, and finally $\epsilon$ is a mean-zero random error. The average treatment effect is the difference in counterfactual scores $ATE = \E[X = 1]{y} - \E[X = 0]{y} = \beta_1$. The OLS estimator for $\beta_1$, $\hat{\beta}_1$, and inference on the hypothesis $H_0: \beta_1 = 0$ reduces to the two-sample t-test for the difference in means.

\textbf{Proof}: The Simple OLS solution for $\hat{\beta}_1 = \frac{\Cov[y]{X}}{\Var{X}} = \frac{\frac{1}{n}\sum_i x_i y_i - \bar{x}\bar{y}}{\frac{1}{n} \sum_i x_i^2 - \bar{x}^2}$ where the sample average $\bar{y}=\frac{1}{n} \sum_i y_i$. Since $x$ is a binary indicator variable, we have $x_i = x_i^2$, $\sum_i x_i = n_T$, and $\bar{x}=n_T/(n_T+n_C)$.
Then $\hat{\beta}_1 = \frac{\sum_i x_i y_i - \bar{y} \sum_i x_i}{\sum_i x_i - \frac{1}{n} (\sum_i x_i)^2} = \frac{\sum_i x_i y_i - \bar{y} \sum_i x_i}{n_T - \frac{1}{n_T + n_C} (n_T)^2} = \frac{\sum_i x_i y_i - \bar{y} \sum_i x_i}{\frac{n_T n_C}{n}}$ where the last equality comes from $n_T - \frac{1}{n_T + n_C} (n_T)^2=n_T\left(\frac{n_T + n_C}{n_T + n_C}-\frac{n_T}{n_T + n_C}\right)=\frac{n_T n_C}{n}$.
At the same time, the ATE from the two-sample t-test can be represented as $\frac{1}{n_T} \sum_i x_i y_i - \frac{1}{n_C} \sum_i (1 - x_i)y_i = \frac{n \sum_i x_i y_i - n_T \sum_i y_i}{n_T n_C} = \frac{\sum_i x_i y_i - \bar{y} \sum_i x_i}{\frac{n_T n_C}{n}}$, but this is $\hat{\beta}_1$ from the regression.

Although many experimentation platforms utilize the two-sample t-test, we show it is equivalent to Simple OLS with counterfactual scoring.
This broader framing of the ATE makes it clear that the analysis of online experiments can utilize regression frameworks and their multiple extensions.

\subsection{Causal Identification in OLS}

While all causal effects are a function of counterfactual scores, not all counterfactual scores are causal effects.
As an example, estimated coefficients from OLS, such as $\hat{\beta}_1$, are not necessarily consistent for their population parameters, such as $\beta_1$. Confounding will prevent the parameter from being identified as a causal effect. 

In order for counterfactual scoring with OLS to produce a consistent estimate for the ATE, we need to rely on a property that makes estimates on a subset of coefficients consistent. Given a model $y = \alpha + \sum_k W_k \beta_{w, k} + \epsilon$, an OLS estimated coefficient $\hat{\beta}_{w, i}$ is consistent for $\beta_{w, i}$ if $\E[W_1, W_2, \ldots]{W_i^T \epsilon} = 0$ and the covariate matrix is full rank (\cite{woolridgeIdentification}). In a randomized and controlled experiment the estimated coefficient on $X$ is consistent by construction: $X$ is randomized so it is uncorrelated with all possible variables. Therefore, in the model $y = \alpha + X\beta_1 + \epsilon$, $\hat{\beta}_1$ is consistent for $\beta_1$ and $\widehat{ATE}$ is consistent for ATE.

\section{Generalizing OLS for Causal Effects Analysis}

So far we have shown equivalence between OLS and the two-sample t-test. We now discuss the several extensions that OLS offers, and motivate a general engineering implementation of it for an experimentation platform. The scalability of this implementation is discussed in section 5.

OLS can take advantage of a general covariate matrix, $M$, in order to reduce the variance of the treatment effect (\cite{gelman}). Suppose the covariate matrix has two types of covariates: the treatment indicator variable, $X$, and other exogenous covariates, $W$, so $M = \begin{bmatrix}
1 & X & W
\end{bmatrix}$ is the matrix concatenation of an intercept, $X$, and $W$ and is size $n \times p$. Then, we assume a linear model $y = \alpha + X{\beta}_1 + W{\beta}_2 + \epsilon.$ Although the model has been extended beyond the t-test equivalent, the ATE on $y$ due to $X$ is identified and can still be computed by scoring the model with two counterfactual inputs: the user has been treated, and the user has not been treated. In both scenarios we hold all other covariates for a user fixed.

\begin{align*}
    ATE &= \E[X = 1, W = W]{y} - \E[X = 0, W = W]{y} \\
        &= \alpha + \beta_1 + \beta_2 \E{W} - \alpha - \beta_2 \E{W} \\
        &= \beta_1
\end{align*}

In this extension, the ATE is estimated using $\widehat{ATE} = \hat{\beta}_1$ and the variance of the estimated ATE is $\Var{\widehat{ATE}} = \Var{\hat{\beta}_1}$.

The above equations are general and simple. Given our model with randomized $X$, we query it twice with two different inputs, then take the difference and average. For linear models we can describe this more generally by expressing the expected change in $y$ is the change in $M$ times $\hat{\beta}$. Then, the average treatment effect is simply the average change in $y$. This gives rise to four fundamental equations for computing the treatment effect, referred to in Definitions 1.

\begin{deffloat}
  \vspace{-1em}
    \begin{align}
      \Delta M &= M^{Treatment} - M^{Control} \\
      &= M(X = 1, W) - M(X = 0, W) \nonumber \\
      \Delta \hat{y} &= \Delta M \hat{\beta} \\
      \widehat{ATE} &= \frac{1}{n} 1_n^T \Delta \hat{y} \\
      K &= \left(\frac{1}{n} 1_n^T \Delta M\right) \nonumber \\
      \Var{\widehat{ATE}} &= K \Cov{\hat{\beta}} K^T
  \end{align}
  \vspace{-1em}
  \caption{The fundamental equations for computing average treatment effects. $M(F=f,\ \textellipsis)$ is the model matrix where feature $F$ is set to have value $f$. $1_n^T$ is a size $n$ row vector of ones.}
\end{deffloat}

We now have a generic way to express the ATE and the variance of the ATE in terms of the regression parameters and the potential outcomes that we want to compare. This is a powerful engineering abstraction because developers of the model can add arbitrary amounts of covariates, and change the form of the model, for example using polynomial expansions, while the logic for the evaluation of the treatment effect remains the same. This leads to a key engineering design principle of separating the fitting process for the model from evaluating the treatment effect.

\subsection{Extending Forms of the Treatment Effect}

We can continue to use our four fundamental equations for even arbitrarily complex linear models.
Suppose the model contains interaction terms, for example $y = \alpha + X\beta_1 + W\beta_2 + (X \cdot W)\beta_3 + \epsilon$, where $(X \cdot W)$ is a feature that interacts $X$ and $W$. The average treatment effect is no longer $\beta_1$. The $\widehat{ATE}$ and its variance are now 
$
    \Var{\widehat{ATE}} = \Var{\hat{\beta}_1} + (\frac{1}{n} \sum_i w_i)^2 \Var{\hat{\beta}_3} + 2 (\frac{1}{n} \sum_i w_i)^2 \Cov[\hat{\beta}_3]{\hat{\beta}_1}
$,
which shows that the treatment effect depends on the value of $w_i$. This can still be written in the general form,
$\widehat{ATE} = \frac{1}{n} 1_n^T \Delta \hat{y}$. The treatment matrix has four column components,
$M^{Treatment} = \begin{bmatrix}
1 & X & W & (X \cdot W)
\end{bmatrix}$.
By setting all $X$ variables to 1 we have
$M^{Treatment} = \begin{bmatrix}
1 & 1 & W & W
\end{bmatrix}$ and setting all $X$ variables to 0 we have
$M^{Control} = \begin{bmatrix}
1 & 0 & W & 0
\end{bmatrix}$. Then $\Delta M = \begin{bmatrix}
0 & 1 & 0 & W
\end{bmatrix}$, and $\widehat{ATE} = \frac{1}{n}1_n^T\Delta M \hat{\beta} = \hat{\beta}_1 + (\frac{1}{n} \sum_i w_i) \hat{\beta}_3$ giving the average treatment effect above. The ATE and the variance on the ATE are computed the same as before in terms of $\Delta M$.

By including the interaction term, we can also query the model for the conditional average treatment effect, CATE. The CATE is the average treatment effect for a specific group of users. In the simplest implementation, we filter the $\Delta y$ vector to the users in the group, and then recompute an average. Let $u$ be a vector of users, and $g$ be a group of users, then
\begin{align*}
    S &= \frac{1}{|g|} (1_n \cdot 1(u \in g)) \\
    \widehat{CATE}(g) &= S^T \Delta y \\
    \Var{\widehat{CATE}(g)} &= (S^T \Delta M) \Cov{\hat{\beta}} (S^T \Delta M)^T.
\end{align*}

Using interaction terms in the model, whether that is to improve the form of the model or to query for heterogeneity in treatment effects, adds little engineering overhead to the evaluation of the treatment effect. Using the general form for linear model treatment effects we do not need to be concerned with how coefficients or covariances should be combined in order to compute the distribution of the average or conditional average treatment effect. This is an engineering simplification as developers can iterate on the form of the model without engineering changes in the XP.

The computation for ATE and CATE are similar where we could design an engineering system that responds to live analysis queries very efficiently. We can fit one model for both ATE and CATE queries, in fact the ATE can be thought of as the CATE computed over all users, $g = U$. When evaluating the treatment effect, the difference between an ATE query and a CATE query is simply the vector multiplier $\frac{1}{|g|} (1_n \cdot 1(u \in g))$. For different CATE queries, we iterate through multiple groups, but that also only creates a small change to $\frac{1}{|g|} (1_n \cdot 1(u \in g))^T$. This insight leads to an engineering simplicity: compute counterfactual scores first, and then aggregate on the fly.

\begin{algorithm}
\caption{Computing Treatment Effects Live}
\begin{algorithmic}[1]

\State Fit a model using $X$, $W$, and $y$.
\State Using the model from step 1, query the the potential outcomes for all users, creating the $\Delta \hat{y}$ vector.
\State Receive an analysis query for ATE or CATE(g).
\State Aggregate the potential outcomes from 2) to compute either ATE or CATE(g).
\State Receive an analysis query for a different $g'$. Without refitting the model, aggregate the potential outcomes again as in step 4.

\end{algorithmic}
\end{algorithm}

In this structure, queries for ATE and CATE can reuse computations from 1 and 2. Although queries for ATE can be done faster by simply looking up the relevant coefficient in a table, standardizing the implementations to use the general form enables one single compute strategy to handle both average and conditional average treatment effects, simplifying the engineering system's design. This is different than the traditional two-sample t-test design of aggregating first, then computing differences in the aggregates.

\subsection{Homoskedastic, Heteroskedastic, and Other Covariance Types}

Introducing interaction terms in the regression allows us to model heterogeneity in treatment effects, where the treatment effect is a function of $X$ and $W$. Likewise, we can make the covariance matrix a function of $X$ and $W$ as well by using heteroskedasticity-robust (HC-robust) covariances (\cite{white1980heteroskedasticity}). Estimating regression coefficients with HC-robust estimated covariances does not change the point estimates of the regression parameters, hence point estimates of the ATE and CATE do not change. However, HC-robust covariances do change $\Cov{\hat{\beta}}$, changing the variance of the treatment effects. We can still have a unified compute strategy by abstracting functions for different types of $\Cov{\hat{\beta}}$, leaving the rest of the linear algebra the same. Furthermore, since point estimates do not change, the structure of our engineering system can still prebuild the $\Delta \hat{y}$ vector, and different variances only get computed during the live query.

\subsection{Time Interactions, Clustered Covariance}

In addition to measuring heterogeneity in treatment effects across groups, we may be interested in measuring heterogeneity across time. To see treatment effects through time, we record multiple observations per user in the experiment, forming a longitudinal study. The treatment effect can vary in time due to the treatment’s diminishing effect or interactions with other experiments.

We may be interested in estimating a regression with user-level fixed effects, time effects, heterogeneous treatment effects, and treatment effects that vary over time. We use the regression below as a leading example:

\begin{align*}
  y_{i,t} &= \beta_0 + \alpha_i + f(t)\beta_t + X_{i,t}\beta_1 + W_{i,t}\beta_2\\
                      &\qquad + (X_{i, t} \cdot W_{i,t})\beta_3 + (X_{i, t} \cdot f(t))\beta_4 + \ldots + \epsilon_{i, t}.
\end{align*}

In this case, the point estimate of the ATE can still be computed by taking the difference in potential outcomes. $\widehat{ATE} = \frac{1}{n} \sum_i \sum_t y_{i, t}(X_{i, t} = 1, W_{i,t} = W_{i,t})-\frac{1}{n} \sum_i \sum_t y_{i, t}(X = 0, W_{i,t} = W_{i, t})$ which can be written in the generalized form where $M$ is a longitudinal matrix. Because the observations in $M$ are not i.i.d., we can use clustered covariances (\cite{clusteredvcov}) to prevent overconfidence in the estimate of the ATE due to autocorrelation within users. Using clustered covariances only changes $\Cov{\hat{\beta}}$; we can be future-proof to longitudinal regression by leaving the variance of the treatment effect in terms of the covariance of the regression parameters.

\section{Simple Engineering Checklist}

We offer a simple engineering checklist in order to establish a good baseline implementation of a simple and generalized system to compute treatment effects using linear models. Afterwards, we offer optimizations to make it more scalable while preserving its generality.

For each outcome:
\begin{todolist}
  \item Construct a model matrix $M$ using the data.
  \item Fit an OLS model using $M$ and $y$ to get $\hat{\beta}$ and $\Cov{\hat{\beta}}$.
  \item Construct two additional model matrices $M^{Treatment}$ and $M^{Control}$ by modifying the model matrix to set everyone in the treatment and control groups, respectively. Construct $\Delta M$ and $\Delta \hat{y} = \Delta M \hat{\beta}$.
  \item Receive a query for either ATE or CATE(g). Filter $\Delta \hat{y}$ according to the relevant users and aggregate.
\end{todolist}

\section{Optimal Numeric Computation for Treatment Effects}

In this section we show how to improve computational efficiency while preserving generalizability.

\subsection{Multiresponse Regression}

Suppose an engineering system is estimating causal effects for the treatment variable $X$ on $m$ different outcomes which are analyzed with the same features: $X$ and $W$. Instead of fitting $m$ OLS models, we fit one multiresponse OLS model. The fitting process for one outcome, $\hat{\beta} = (M^T M)^{-1} M^T y$, can be extended when all models use the same $M$ matrix by replacing the matrix-vector multiplication $M^T y$ with the matrix-matrix multiplication $M^T Y$, where $Y$ is the concatenation of all outcome vectors. $\hat{\beta}$ will return a matrix of coefficients that has dimensions $p \times m$ and is the concatenation of all estimated coefficients. Likewise, to compute the covariance matrices, we compute and save $(M^T M)^{-1}$ exactly once for all $m$ outcomes. Finally, we need the estimated residuals, which is computed in a vectorized way as $\hat{\epsilon} = Y - M\hat{\beta}$.

\subsection{Use Sparse Data Structures}

The model matrix, $M$ can have many zeros in it. In practice, the dummy variable encoding of a categorical variable (\cite{dummyvariable}) in the data will already make $M$ sparse. The storage can be optimized by using sparse matrix data structures, for example the compressed sparse column format. Furthermore, the linear algebra to estimate OLS can be accelerated using sparse linear algebra. For example, the multiplication $M^T M$ in OLS normally runs $np^2$ floating point operations, but if the sparsity rate of $M$ is $s$, then an optimized sparse linear algebra library only needs to compute $(n \cdot s)p^2$ operations. $M^T M$ where $M$ has categorical variables with many levels also tends to be sparse, so it can be inverted using a sparse Cholesky decomposition. Golub and Van Loan's \textit{Matrix Computations} (\cite{golub}) shares many matrix factorization algorithms to solve large sparse linear systems efficiently. Implementations are also found in numerical libraries such as Eigen (\cite{eigenweb}).

\subsection{Compress Data to Sufficient Statistics}

For statistical tests using distributions from the exponential family (\cite{efron1978geometry, hipp1974sufficient}), we compress the data to the sufficient statistics, which by definition are the minimal information needed to recover the distribution of the data. OLS is based on the normal distribution, which has sufficient statistics $n$, $\sum y$, and $\sum y^2$. \cite{compression} show how to adapt the fitting process for OLS to use sufficient statistics, and how to get good compression rates without biasing the treatment effect.

\subsection{Aggregating $\Delta \hat{y}$ for ATE, CATE}

In the four fundamental equations, we described aggregating the $\Delta \hat{y} = \Delta M \hat{\beta}$ vector to estimate the treatment effect. For 0-1 treatment variables it is not necessary to materialize $\Delta \hat{y}$; in fact it is an extremely expensive operation. Since $\Delta \hat{y}$ will be aggregated, we can directly compute the necessary parts for that aggregation.

For $\widehat{ATE} = \frac{1}{n} 1_n^T \Delta \hat{y} = \frac{1}{n} 1_n^T \Delta M \hat{\beta}$, we first isolate $\frac{1}{n} 1_n^T \Delta M$ by taking the column means of $\Delta M$, yielding a length $p$ row vector. If the model is in the form $y = \alpha + X\beta_1 + W\beta_2$ then $\Delta M = \begin{bmatrix}
0 & 1 & 0
\end{bmatrix}$ and the column means of $\Delta M$ are
$\frac{1}{n} 1_n^T \Delta M = \begin{bmatrix}
0 & 1 & 0
\end{bmatrix}$. This vector can be constructed without materializing any copies of the model matrix. By multiplying this with $\hat{\beta}$ the result will isolate the $\hat{\beta}_1$ coefficient specifically, which is the $\widehat{ATE}$.

In the case of CATE(g), we compute the average treatment effect for a particular group, $g$. For each $g$, $\widehat{CATE}(g) = \frac{1}{|g|} 1_{|g|}^T \Delta M_g \hat{\beta}$, where $\Delta M_g$ is the submatrix of $\Delta M$ containing the rows pertaining to users in group $g$.
If the model is in the form $y = \alpha + X\beta_1 + W\beta_2 + (X \cdot W)\beta_3$ then $\Delta M = \begin{bmatrix}
0 & 1 & 0 & W
\end{bmatrix}$ and the column means of $\Delta M_g$ are
$\frac{1}{|g|} 1_{|g|}^T \Delta M_g = \begin{bmatrix}
0 & 1 & 0 & \bar{W}_g
\end{bmatrix}$.
This is a fast operation since it does not require manipulating the data and constructing counterfactual matrices. When fitting the OLS model, a model matrix, $M$, was already constructed which can be partitioned by $g$ and used to directly compute $\bar{W}_g$.

The variance of the treatment effect is also computed very efficiently from the column means of $\Delta M$. No counterfactual model matrices are needed.

\subsection{Sorting Optimizes CATE}

From optimization 5.4, when computing $CATE(g)$ we need to be able to identify the relevant submatrix in $\Delta M$ to aggregate, which reduces to finding the row numbers in $M$ for users in $g$. When querying for multiple groups, the search for the users can be expensive. To optimize this we can sort the data according to the grouping variables $g_1, g_2, \ldots$, which will allow us to find relevant users for group $g$ quickly.

By sorting we pay an upfront cost of $O(nlog(n))$. After the data has been sorted, users for a group $g$ are in consecutive rows. Using a single pass over the sorted data we identify the relevant submatrices of $\Delta M$ to aggregate for any group quickly. Suppose there are $G$ groups we wish to compute the CATE for. The cost with sorting is $O(nlog(n) + n)$. Without sorting, each group needs to scan the entire dataset to find relevant users and the complexity is $O(nG)$. When $G$ is sufficiently large, the gain from forming consecutive rows outweighs the cost of sorting.

\subsection{Optimized Engineering Checklist}

We offer a second checklist for computing linear model treatment effects efficiently.
Given many outcomes, identify outcomes that have the same regression form and create an outcome group. For each outcome group:

\begin{todolist}
  \item Construct one sparse model matrix $M$ using the data.
  \item Compress the data by grouping rows of the $M$ matrix and computing the weights, $\sum y$, and $\sum y^2$ for the corresponding rows in $Y$.
  \item Fit a sparse-optimized weighted OLS model using $M$ and $Y$ to get $\hat{\beta}$ and $\Cov{\hat{\beta}}$. Adjust $\Cov{\hat{\beta}}$ for compression using the weights, $\sum y$, and $\sum y^2$.
  \item Receive a query for either ATE or CATE. If ATE, compute the column means of $\Delta M$ and multiply with $\hat{\beta}$. Use the column means of $\Delta M$ again to compute $(\frac{1}{n} 1_n^T \Delta M) \Cov{\hat{\beta}} (\frac{1}{n} 1_n^T \Delta M)^T$. If the query is for CATE with grouping variables $g_1, g_2, \ldots$, sort $M$ according to the grouping variables, then partition the row numbers of the sorted $M$ by group. For each group, filter the sorted $M$ to the users in $g$ and compute the conditional treatment effect as if it were the average treatment effect.
\end{todolist}

\section{Performance Benchmarks}

We now demonstrate the performance of an optimized treatment effect estimator using OLS. The purpose of this section is to show that large linear models, beyond the subclasses of the two-sample t-test, CUPED, and FORME, can be made performant enough to deploy to a production-level web service, as well as be run interactively in a local environment. The performance benchmarks use data for an experiment with eight different treatment policies, 10 metrics, and a sparse set of covariates. We vary the size of the experiment from a sample size of $10^5$ to $5\cdot10^7$ users, and measure the time to fit the model as well as evaluate the treatment effect. The largest experiment computes 70 ATEs in 10 seconds, and 700 CATEs in 15 seconds. The computation for the CATEs can also be reused to compute another set of CATEs on a different group efficiently.

\begin{figure}[h]
\includegraphics[width=\columnwidth]{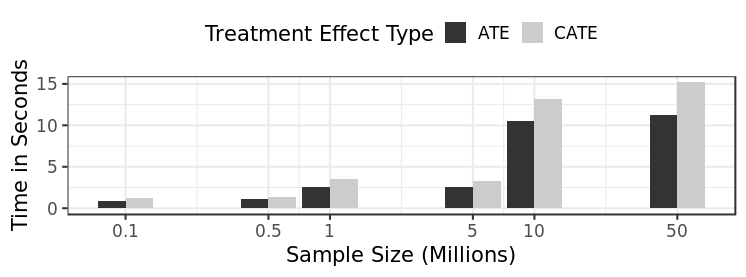} 
\caption{Time to compute treatment effects.}
\end{figure}

Using the optimized computational strategy, the OLS estimator for the ATE and CATE can be scaled to be fast enough to run on a single machine. This is the greatest simplification to an XP as well as causal inference researchers, because it reduces the need and overhead for a distributed computing environment.

\section{Conclusion}

Linear models are fundamental to causal inference and should be accessible to an experimentation platform; they provide a general framework to estimate average, conditional average, and dynamic treatment effects. An engineering system that uses linear models needs to be future-proof to new forms of the model and needs a simple, yet comprehensive, architecture. Using a counterfactual scoring framework with OLS, we can create a computational strategy for these effects. All of these variations can be estimated using the framework: $\Delta \hat{y} = \Delta M \hat{\beta}$. For many use cases, we can compute the difference in counterfactual outcomes without constructing counterfactual model matrices, a great simplification and optimization. By combining several other optimizations, we make the engineering system scalable and practical for online use cases. Simultaneously, the highly performant linear model estimators can exist without distributed systems, and can be used locally by other scientists and developers.

\printbibliography

\end{document}